\documentclass[prd,superscriptaddress,onecolumn,showkeys]{revtex4}
\usepackage{textcomp}
\usepackage{eurosym}
\usepackage{amsfonts}
\usepackage{array}
\usepackage{amsthm}
\usepackage{bm}
\usepackage{palatino}
\usepackage[colorinlistoftodos]{todonotes}
\usepackage{mathpazo}
\usepackage{supertabular}
\usepackage{subfig}
\usepackage{amssymb}
\usepackage{eurosym}
\usepackage{amsmath}
\usepackage{epsfig}
\usepackage{graphics}
\usepackage{changes}
\usepackage{color}
\usepackage{graphicx}
\usepackage[colorlinks=true,
            linkcolor=blue,
           urlcolor=black,
           citecolor=black]{hyperref}

\def\be{\begin{equation}}
\def\ee{\end{equation}}
\def\beq{\begin{eqnarray}}
\def\eeq{\end{eqnarray}}

\def\bes{\begin{eqnarray}}
\def\ees{\end{eqnarray}}

\begin{document}
\title{Tunneling and entropy analysis of parameterized black hole with rotating case}
\author{Muhammad Asgher}
\email{m.asgher145@gmail.com}
\affiliation{Department of Mathematics, The Islamia
University of Bahawalpur, Bahawalpur-63100, Pakistan}

\author{Anosha Karamat}
\email{anosha.karamat2@gmail.com}
\affiliation{Department of Mathematics, Minhaj University, Lahore-54000, Pakistan}

\author{Rimsha Babar}
\email{rimsha.babar10@gmail.com}
\affiliation{Division of Science and Technology, University of Education, Township, Lahore-54590, Pakistan}

\author{Riasat Ali$^{*}$}
\email{riasatyasin@gmail.com}
\affiliation{Department of Mathematics, GC,
University Faisalabad Layyah Campus, Layyah-31200, Pakistan}

\begin{abstract}
In this work, we study the parameterized black hole
solution by applying the Newman-Janis approach and also examine the Hawking temperature. We consider a Lagrangian field equation associated with the generalized uncertainty principle to study the motion of boson particles. By using semi-classical phenomenon, we analyze the modified Hawking temperature and graphically check the effects of deformation, rotation and correction parameter on black hole geometry. Furthermore, we investigate the logarithmic
corrected entropy and also analyze the graphical behavior of deformation and quantum gravity parameter on the
logarithmic corrected entropy of black hole.
\end{abstract}


\date{\today}

\maketitle

\section{Introduction}
Black Holes (BHs) are a very dominating objects in the field of quantum gravity (QG). Many types of BH solutions in different theories have been investigated. Alternately, to find the specific BHs solutions as representation in gravitational theory, it is interesting to use a different approach with parametric solutions \cite{1}. Many parameterized BH solutions were presented and examined in the recent research with various inspirations \cite{2}-\cite{5}. Chrusciel and his colleagues \cite{6} have investigated the Konoplya–Zhidenko parameterization with a static deformation that allows the resulting metric to violates the no-hair theorem. Cardoso et. al., \cite{4} have studied the parameterized BH in quasi-normal modes and analyzed decoupled equations for non-rotating BHs. Magalhães and his fellows \cite{8} have studied the parameterized BH with a single deformation parameter and investigated the impacts of deformation parameter on null geodesics. Konoplya and Zhidenko \cite{9} have computed the general parameterization of a BH metric in various gravity theories with many parameters i.e., parameters of shadow, electromagnetic radiation, quasi-normal modes and accretion matter.

The interesting class of charged rotating BHs are depicted by Kerr-Newman metrics,
that can be created by a procedure known as the Newman-Janis strategy \cite{7}. Whereas this strategy is especially effective for inferring the Kerr metric and its electromagnetic speculation, the Kerr-Newman spacetime, it has frequently been criticized \cite{10} because the
system is definitely not an overall strategy for creating vacuum from vacuum metrics as well as there is a sure discretion in the decision of complexification of terms in the unique basic metrics (Schwarzschild or Reissner Nordstr$\ddot{o}$m). The Kerr space-time portrays the metric of a rotating BH, it is actually of interest whether it depicts the outside of a modified
axisymmetric rotating gigantic object. Whereas Birkhoff’s theorem \cite{11} states that the metric outside to a spherically symmetric gigantic object is uniquely expressed by the Schwarzschild metric there's shockingly no reason to expect that the metric outside to an arbitrary stationary axisymmetric perfect fluid ought to be the Kerr metric. Although,  stays an open problem whether the Kerr arrangement can performs the outside of any perfect fluid by any means. The metrics for different types of BHs and black rings for rotating and non-rotating cases have been widely discussed in literature \cite{13, 13a, 15a}.

Moreover, to discuss the thermodynamics of BHs is always a very interesting feature for researchers. In this regard, many approaches has been used and the quantum tunneling method is also a very popular technique to analyze the thermodynamics like Hawking temperature of BHs. It is also a well known fact that quantum gravity influences the physical properties of gigantic objects. In order to reduce the gravitational effects, researchers take help of generalized uncertainty principle (GUP). The GUP is a sufficient tool for depicting BHs and their thermodynamic properties. The researchers calculate the corrected temperature and entropy with the help of GUP. The GUP is associated to the Heisenberg principle \cite{17, 18}.
Ali \cite{19} has discussed the formation of  mini BHs via GUP. Faizal and Khalil \cite{20} have introduced the universality law of entropy related to its area and discussed the quantum corrections of thermodynamics of black objects.  Gangopadhyay and Dutta \cite{21} have investigated the corrections to the thermodynamics of a Schwarzschild BH for higher dimensions incorporating GUP. Gecim, and Sucu \cite{22} have studied the GUP effects for fermions and scalar particles by utilizing the Hamilton–Jacobi technique for Martinez–Zanelli BH and calculated the tunneling probability and temperature for the corresponding BH. They have also investigated the modified temperature of BTZ BH by using the quantum tunneling approach for scalar particles \cite{23}. The quantum corrected temperature for different kind of BHs through tunneling method based on GUP has been discussed in detail \cite{28}. Moreover, by using the quantum tunneling phenomenon for vector particles, the Hawking temperature for BHs have been anaylzed \cite{29, 30,31, 32, 33, c4}. Hassanabadi and his colleagues \cite{34} have analyzed a detailed thermodynamics analysis for a charged BH via GUP.

The important motivation of this work is to introduce the rotating case of parameterized BH through Newman-Jannis approach as well as to derive the thermodynamics of parameterized BH that add the spin parameter in the metric. Moreover, to discuss the effects of rotation parameter in the presence of quantum gravity.

We have formed our paper as follows; In Sec. \textbf{II},
we derive a Parameterized BH solution in the context of Newman-Janis algorithmic rule and also investigate
the Hawking temperature for the corresponding BH. In Sec. \textbf{III}, we investigate the graphical importance of temperature under the effects of deformation parameter as well as rotation parameter. In Sec. \textbf{IV}, we examine the modified tunneling and modified temperature. Sec \textbf{V}, gives the analysis of Hawking temperature with respect to horizon and check the stabile state of parameterized BH with rotation parameter under quantum gravity effects. Sec \textbf{VI}, studies the logarithmic entropy corrections for parameterized BH in the rotating case. In Sec. \textbf{VII}, we analyze the physical significance of corrected entropy plots under the effects of deformation and correction parameter for rotating parameterized BH.
Sec \textbf{VIII}, we have analyzed our results.

\section{Parameterized Black Hole With Rotation Parameter}

We consider the parametrization studied in Ref. \cite{P1}, which the BH solution modifies into extra parameters incorporating in the mass like term known as deformation parameters. The non-spinning form's parametrized like BH solution is provided \cite{P2} by
\begin{equation}
ds^{2}=-A_{KZ}(r)dt^2+\frac{1}{A_{KZ}}(r)dr^2+r^{2}d\Omega^2,
\end{equation}
with $A_{KZ}(r)=1-\frac{2M(r)}{r}$.
In order to include more parameters, the mass-term $M(r)$ is used as \begin{equation}
2M(r)=\sum_{i=0}^{\infty}\frac{\eta_{i}}{r_{i}}+2M,
\end{equation}
with $\eta_{i}$ represent the deformation parameters. When $\eta_{i}=0$,  we get Schwarzschild BH solution. In the BH family of parametrized solutions, a group of solutions with the same Schwarzschild post-Newtonian asymptotics can be chosen.
In this section, we analyze the parameterized BH solution in the frame of
Newman-Janis approach. Additionally, we calculate the Hawking temperature for the parameterized BH metric. To do so, we choose a static spherically symmetric BH metric \cite{8}
\begin{equation}
ds^{2}=-A(r)dt^2+\frac{1}{A(r)}dr^2+r^2 d\theta^2+r^2 \sin^2\theta d\phi^2,\label{sol}
\end{equation}
where
\begin{equation}
A(r)=1-\frac{2M}{r}-\frac{\eta}{r^{3}},\nonumber
\end{equation}
here, $M$ and $\eta$ represents the mass and deformation parameters of BH, respectively.

To study the solution of parameterized BH in the frame of Newman-Janis approach.
Firstly, we consider a transformation in the background of parameterized BH in Eq. (\ref{sol}) from the
coordinates $(t, r, \theta, \phi)$ to new coordinates $(w, r, \theta, \phi)$ as
\begin{eqnarray}
dw=dt-\frac{dr}{A(r)},\label{A}
\end{eqnarray}
the metric Eq. (\ref{sol}) can be rewritten in the given form under the
defined transformation
\begin{equation}
ds^{2}=-A(r)dw^2-2dwdr+r^2 d\theta^2+r^2 \sin^2\theta d\phi^2.
\end{equation}
The inverse metric elements are defined in the form
\begin{equation}
g^{01}=g^{10}=-1,~~g^{11}=A,~~g^{22}=\frac{1}{r^2},~~g^{33}=\frac{1}{r^2\sin^2\theta}.
\end{equation}

The inverse components of the metric in the form of null tetrad are represented as
\begin{eqnarray}
g^{ij}=-l^j n^i-l^i n^j+m^i \bar{m}^{j}+m^j \bar{m}^{i}.
\end{eqnarray}
The elements related to null tetrad can be given as
\begin{eqnarray}
l^{i}&=&\delta_{r}^{i},~~~n^{i}=\delta_{w}^{i}-\frac{1}{2} A \delta_{r}^{i},\nonumber\\
m^{i}&=&\frac{1}{\sqrt{2}r} \delta_{\theta}^{i}+\frac{\iota}{\sqrt{2}r \sin \theta}\delta_{\phi}^{i},\nonumber\\
\bar{m}^{i}&=&\frac{1}{\sqrt{2}r} \delta_{\theta}^{i}-\frac{\iota}{\sqrt{2}r \sin \theta}\delta_{\phi}^{i}.\nonumber
\end{eqnarray}
Although, at the some points in the parameterized BH metric, the relationship of null tetrad with vectors is defined as
\begin{equation}
l_{i}l^{i}=n_{i}n^{i}=m_{i}m^{i}=l_{i}m^{i}=m_{i}m^{i}=0,\nonumber
\end{equation}
and
\begin{equation}
l_{i}n^{i}=-m_{i}\bar{m}^{i}=1.\nonumber
\end{equation}
The transformation of coordinates in the $(w, r)$ plane can be expressed as
\begin{eqnarray}
w&\rightarrow &w-\iota a\cos\theta,\nonumber\\
r&\rightarrow &r+\iota a\cos\theta.\nonumber
\end{eqnarray}
Furthermore, we compute the transformations in the following form
\begin{equation}
A(r)\rightarrow \tilde{A}(r, a, \theta),
\end{equation}
and
\begin{equation}
r^2+a^2 \cos^2\theta=\sigma^2.
\end{equation}
 The null vectors in the $(w, r)$ plane obtain the following form
\begin{eqnarray}
l^{i}&=&\delta_{r}^{i},~~~n^{i}=\delta_{u}^{i}-\frac{1}{2}
\tilde{A}\delta_{r}^{i},\nonumber\\
m^{i}&=&\frac{1}{\sqrt{2}r}\left(\delta_{\theta}^{i}+\iota a \sin\theta(\delta_{u}^{i}
-\delta_{r}^{i})+\frac{\iota}{\sin\theta}\delta_{\phi}^{\iota}\right),\nonumber\\
\bar{m}^{i}&=&\frac{1}{\sqrt{2}r}\left(\delta_{\theta}^{i}-\iota a \sin\theta(\delta_{u}^{i}
-\delta_{r}^{i})-\frac{\iota}{\sin\theta}\delta_{\phi}^{i}\right).
\end{eqnarray}
By the definition of null tetrad, the significant inverse elements of the metric
$g^{i j}$ in the coordinates $(w, r, \theta, \phi)$ can be computed in the form
\begin{eqnarray}
g^{00}&=&\frac{a^2 \sin^2\theta}{\sigma^2},~~~g^{01}=g^{10}=-1-\frac{a^2\sin^2\theta}{\sigma^2},
~~~g^{11}=\tilde{A}(r, \theta)+\frac{a^2\sin^2\theta}{\sigma^2},~~~
\nonumber\\
g^{33}&=&\frac{1}{\sigma^2\sin^2\theta},~~~g^{03}=g^{30}=\frac{a}{\sigma^2},~~~
g^{13}=g^{31}=-\frac{a}{\sigma^2},~~~g^{22}=\frac{1}{\sigma^2},\nonumber
\end{eqnarray}
here
\begin{equation}
\tilde{A}(r, \theta)=\frac{r^2 A+a^2 \cos^2\theta}{\sigma^2}.
\end{equation}
Moreover, we derive a transformation of coordinates from $(w, r, \theta, \phi)$
to the  coordinates $(t, r, \theta, \phi)$ in the following state
\begin{equation}
dw=dt+\zeta(r)dr,~~~d\phi=d\phi+\xi(r)dr\label{lam},
\end{equation}
here \begin{eqnarray}
\zeta(r)&=-&\frac{r^2+a^2}{r^2 A +a^2},\nonumber\\
 \xi(r)&=&-\frac{a}{r^2 A +a^2}.\nonumber
\end{eqnarray}
At last, we obtain the BH solution in the given condition
\begin{equation}
ds^{2}=-\tilde{A}(r)dt^2+\sigma^2 d\theta^2+\frac{\sigma^2}{\tilde{A}(r)}dr^2+2a \sin^2\theta\Big[\tilde{A}(r)-1\Big]dt d\phi
+\sin^2\theta\Big[a^2\sin^2\theta\Big(2-\tilde{A}(r)\Big)+\sigma^2\Big]d\phi^2.
\end{equation}
The above equation can be derived in the given form
\begin{eqnarray}\label{ds}
ds^{2}&=&-\left(1-\frac{2Mr+\eta r^{-1}}{\sigma^2 }\right)dt^2+\frac{\sigma^2 }{\Delta_{r}}dr^2-2a \sin^2\theta\left(\frac{2Mr+\eta r^{-1}}{\sigma^2}\right)dt d\phi+\sigma^2 d\theta^2\nonumber\\
&+&\sin^2\theta\Big[a^2\sin^2\theta\Big(\frac{2Mr+\eta r^{-1}}{\sigma^2}\Big)+\sigma^2\Big]d\phi^2.
\end{eqnarray}
here
\begin{equation}
\Delta_{r}=r^2 -2Mr+a^{2}-\eta r^{-1}.
\end{equation}
In order to study the thermodynamics of the corresponding BH, we derive the temperature in the presence of rotation parameter with the help of surface gravity $\kappa$ by using the following formula
\begin{equation}
    T_{H}=\frac{\kappa}{2\pi},
\end{equation}
where
\begin{equation}
   \kappa=\frac{\Delta'_{r}}{2(r^{2}_{+}+a^{2})},~~~~~\text{and}~~~~~\Delta'_{r}=\frac{d}{dr}(\Delta_{r}).
\end{equation}
The Hawking temperature for parameterized BH with the rotation parameter can be derived as
\begin{equation}\label{2}
T_{H}=\frac{2Mr_{+}^4+3\eta r_{+}^2+a^2 (\eta-2Mr_{+}^2)}{4\pi r_{+}^{2}\Big(r_{+}^2 +a^2\Big)^2}.
\end{equation}
The temperature $T_{H}$ depends on spin parameter $a$, BH mass $M$ and deformation parameter $\eta$.
It is very important to express here that in the absence of spin parameter i.e., $a=0$, we get the Hawking temperature which is independent
of Newman-Janis algorithmic rule.

The temperature in its original form can be obtained in the following form
\begin{equation}
T_{H}=\frac{2Mr_{+}^4+3\eta r_{+}^2}{4\pi r_{+}^6}.
\end{equation}
The above expression indicates the temperature of parameterized BH in the absence of rotation parameter.

\section{Graphical interpretation of Temperature versus horizon}
This section studies the graphical conduct of $T_{H}$
via horizon. We analyze the physical
understanding of the plots with the impacts of deformation and rotation parameters.
We discuss the stable conditions of parameterized BH in the rotating case.

\begin{center}
\includegraphics[width=7cm]{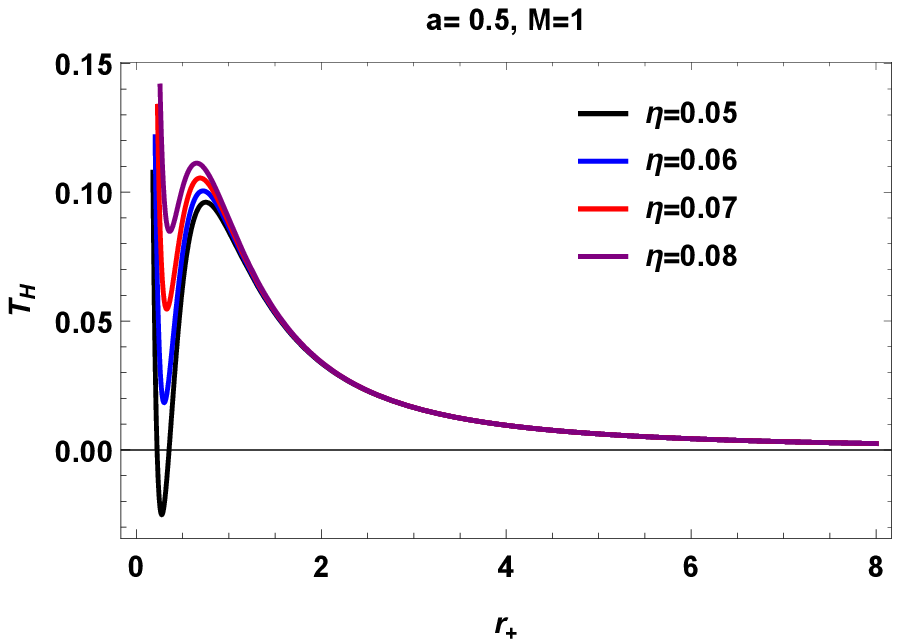}\includegraphics[width=7cm]{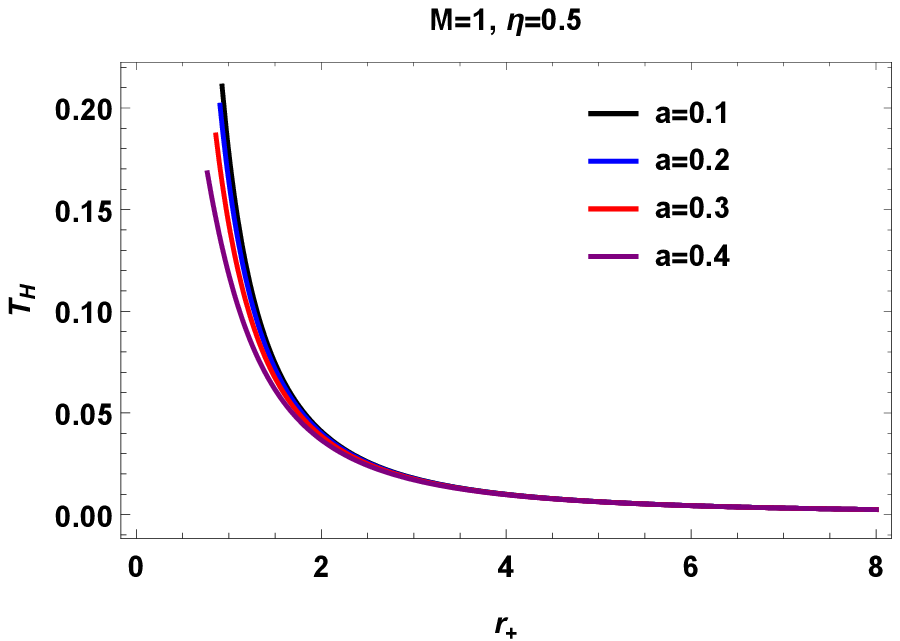}\\
\textbf{Fig. 1}: Temperature versus horizon for $M=1$. Left $a=0.5$, $\eta=0.05$ (black), $\eta=0.06$ (blue), $\eta=0.07$ (red), and $\eta=0.08$ (purple). Right $\eta=0.06$, $a=0.1$ (black), $a=0.2$ (blue), $a=0.3$ (red), and $a=0.4$ (purple).\end{center}

The left plot in \textbf{Fig. 1} shows that the $T_{H}$ constantly decreases and after a minima it rises to reach at a maxima and then again gradually decrease to shows an asymptotically flat description until $r\rightarrow\infty$. There can be seen two critical values $r_{+}^{min}$ and $r_{+}^{max}$ in the plot where the BH is stable for $r_{+}^{min} \leq r_{+}\leq r_{+}^{max}$. The $T_{H}$ rises with the rising values of deformation parameter $\eta$.

The right plot gives that the $T_{H}$ constantly decreases and then approaches to an asymptotically flat behavior upto $r_{+}\rightarrow \infty$ in the range $0\leq r_{+}\leq8$. Moreover, for the rising values of $a$, the $T_{H}$ shows a decreasing conduct.

\section{Modified Tunneling of Parameterized Black Hole with Rotation Parameter}
In this chapter, we examine the quantum mechanical process of Hawking
temperature as a tunneling process for the parameterized BH with rotation parameter. We
compute the tunneling rate of the charged boson and compute the
corrected temperature for parameterized BH. From the Eq. (\ref{ds}), we get
\begin{equation}\label{dv}
ds^{2}=-\Delta dt^2+\frac{1}{\Delta}dr^2+\sigma^2 d\theta^{2}
+U d\phi^{2}+2Vdt d\phi,
\end{equation}
with $\Delta=\frac{\sigma^2 r}{r^2 -2Mr+a^{2}-\eta r^{-1}},~~ U=\sin^2\theta\Big[\sigma^2+a^2\sin^2\theta\Big(\frac{2Mr+\eta r^{-1}}{\sigma^2}\Big)\Big]$ and $V=-a \sin^2\theta\left(\frac{2Mr^2+\eta}{\sigma^2 r}\right).$\\

Now, we choose the emission of boson particles from
parameterized BH with rotation parameter in the background of tunneling process around the horizon.
This is done by solving charged Lagrangian equation with GUP parameter for vector field $\psi$, which is given by \cite{15a}
\begin{equation}
\partial_{i}(\sqrt{-g}\psi^{ji})+\sqrt{-g}\frac{m^2}{\hbar^2}\psi^{j}+\sqrt{-g}\frac{\iota}{\hbar}A_{i}\psi^{ji}
+\sqrt{-g}\frac{\iota}{\hbar}eF^{ji}\psi_{i}+\hbar^{2}\beta\partial_{0}\partial_{0}\partial_{0}(\sqrt{-g}g^{00}\psi^{0j})
-\hbar^{2}\beta \partial_{i}\partial_{i}\partial_{i}(\sqrt{-g}g^{ii}\psi^{ij})=0,\label{L}
\end{equation}
here $\beta$ is the GUP or quantum gravity parameter, $g$ denotes the determinant of the metric, $\psi^{ji}$ denotes the anti-symmetric tensor as well as $m$ stands for mass of particle, since
\begin{equation}
\psi_{ji}=(1-\beta{\hbar^2\partial_{j}^2})\partial_{j}\psi_{i}-
(1-\beta{\hbar^2\partial_{i}^2})\partial_{i}\psi_{j}+
(1-\beta{\hbar^2\partial_{j}^2})\frac{\iota}{\hbar}eA_{j}\psi_{i}
-(1-\beta{\hbar^2}\partial_{\nu}^2)\frac{\iota}{\hbar}eA_{i}\psi_{j},~~~
F_{ji}=\nabla_{j} A_{i}-\nabla_{i} A_{j},\nonumber
\end{equation}
here $\beta,~A_{i},~e~$ and $\nabla_{i}$ shows the dimensionless positive parameter,
vector potential of parameterized BH, the particle's charge and covariant derivatives, respectively.
The non zero values of $\psi^{i}$ and $\psi^{ji}$ are computed as
\begin{eqnarray}
&&\psi^{0}=\frac{-U\psi_{0}+V\psi_{3}}{U\Delta+V^2},~~~\psi^{1}=\Delta\psi_{1},
~~~\psi^{2}=\frac{\psi_{2}}{\sigma^2},~~~
\psi^{3}=\frac{V\psi_{0}+\Delta\psi_{3}}{U\Delta+V^2},~~\psi^{12}=\frac{\Delta \psi_{12}}{\sigma^2},
~~\psi^{13}=\frac{\Delta \psi_{13}}{U\Delta+V^2},\nonumber\\
&&\psi^{01}=\frac{-U\psi_{01}+V\psi_{13}}{(U\Delta+V^2)},~~~
\psi^{02}=\frac{-U\psi_{02}}{\sigma^2(U\Delta+V^2)},
~~~\psi^{03}=\frac{(U\Delta+\Delta^2)\psi_{03}}{(U\Delta+V^2)^2},~
\psi^{23}=\frac{V\psi_{02}+\Delta\psi_{23}}{\sigma^2(U\Delta+V^2)}.
\end{eqnarray}
The WKB method \cite{13} for vector particles can be defined in the way
\begin{equation}
\psi_{j}=c_{j}\exp\Big[\frac{\iota}{\hbar}I_{0}(t,r, \theta,  \phi,)+
\Sigma \hbar^{n}I_{n}(t,r, \theta,  \phi,)\Big].\label{wkb}
\end{equation}
By putting the Eqs. (\ref{dv}) and (\ref{wkb}) into Eq. (\ref{L}), we collect a set of field equations in \textbf{APPENDIX A} and after applying
the technique of separation variables \cite{15a}, we can get
\begin{equation}
I_{0}=-(E-j\omega)t+W(r)+J\phi+\nu(\theta),
\end{equation}
where $E$ denotes the particle's energy, $J$ represents the particle's angular
momentum associated with angle $\phi$. After substituting the separation of variables into set of field equations, we obtain
\begin{equation*}
Z(c_{0},c_{1},c_{2},c_{3})^{T}=0,
\end{equation*}
which implies a $4\times4$ matrix labeled as "$Z$" in \textbf{APPENDIX B}, whose components
are given as follows
\begin{eqnarray}
Z_{00}&=&\frac{{-U\Delta}}{(U\Delta+V^2)}[W_{1}^2+\beta W_{1}^4]-\frac{U}{\sigma^2(U\Delta+V^2)}[J^2+\beta J^4]
-\frac{U\Delta}{(U\Delta+V^2)^2}[\nu_{1}^2+
\beta \nu_{1}^4]-\frac{m^2 U}{(U\Delta+V^2)},\nonumber\\
Z_{01}&=&\frac{-D\Delta}{(U\Delta+V^2)}[(E-j\omega)+\beta (E-j\omega)^3+eA_{0}+\beta eA_{0}(E-j\omega)^2]W_{1}
+\frac{V\Delta}{(U\Delta+V^2)}+[\nu_{1}+\beta \nu_{1}^3],\nonumber\\
Z_{02}&=&\frac{-U\Delta}{\sigma^2(U\Delta+V^2)}[(E-j\omega)+\beta (E-j\omega)^3-eA_{0}-\beta eA_{0}(E-j\omega)^2]J,\nonumber\\
Z_{03}&=&\frac{-V\Delta}{(U\Delta+V^2)}[W_{1}^2+\beta W_{1}^4]- \frac{U\Delta}{\sigma^2(U\Delta+V^2)^2}[(E-j\omega)+\beta (E-j\omega)^3
-eA_{0}-\beta eA_{0}(E-j\omega)^2]\nu_{1}+\frac{m^2V}{(U\Delta+V^2)^2},\nonumber\\
Z_{10}&=&\frac{-U\Delta}{(U\Delta+V^2)}[(E-j\omega)W_{1}+\beta (E-j\omega)W_{1}^3]-m^2\Delta
-\frac{eA_{0}U\Delta}{(U\Delta+V^2)}[W_{1}+\beta W_{1}^3],\nonumber\\
Z_{11}&=&\frac{-U\Delta}{(U\Delta+V^2)}[(E-j\omega)^2+\beta (E-j\omega)^4-eA_{0}(E-j\omega)-\beta eA_{0}(E-j\omega)W_{1}^2]
+\frac{V\Delta}{(U\Delta+V^2)}+[\nu_{1}+
\beta \nu_{1}^3](E-j\omega)\nonumber\\
&-&\frac{\Delta}{\sigma^2}[J^2+\beta J^4]-
\frac{\Delta}{(U\Delta+V^2)}[\nu_{1}+\beta \nu_{1}^3]
-m^2\Delta-\frac{eA_{0}U\Delta}{(U\Delta+V^2)}
[(E-j\omega)+\beta (E-j\omega)^3-eA_{0}-\beta eA_{0}(E-j\omega)^2]\nonumber\\
&+&\frac{eA_{0}V\Delta}{(U\Delta+V^2)}[\nu_{1}+
\beta \nu_{1}^3],\nonumber\\
Z_{12}&=&\frac{\Delta}{\sigma^2}[W_{1}+\beta W_{1}^3]J,\nonumber\\
Z_{13}&=&\frac{-V\Delta}{(U\Delta+V^2)}[W_{1}+\beta W_{1}^3](E-j\omega)+ \frac{\Delta}{(U\Delta+V^2)^2}[W_{1}+\beta W_{1}^3]\nu_{1}
+\frac{V\Delta eA_{0}}{(U\Delta+V^2)}[W_{1}+\beta W_{1}^3],\nonumber\\
Z_{20}&=&\frac{U}{\sigma^2(U\Delta+V^2)}[(E-j\omega)J+\beta (E-j\omega)J^3]+
\frac{V}{\sigma^2(U\Delta+V^2)}[(E-j\omega)+\beta (E-j\omega)^3\nu_{1}^2]
-\frac{UeA_{0}}{\sigma^2(U\Delta+V^2)}[J+\beta J^3],\nonumber\\
Z_{21}&=&\frac{\Delta}{\sigma^2}[J+\beta J^3]W_{1},\nonumber\\
Z_{22}&=&\frac{U}{\sigma^2(U\Delta+V^2)}[(E-j\omega)^2
+\beta (E-j\omega)^4-eA_{0}(E-j\omega)-\beta eA_{0}(E-j\omega)]
-\frac{\Delta}{\sigma^2}+\frac{V}{\sigma^2(U\Delta+V^2)}[(E-j\omega)
\nonumber\\
&+&\beta (E-j\omega)^3-eA_{0}-\beta eA_{0}(E-j\omega)^2]\nu_{1}-\frac{\Delta}{\sigma^2(U\Delta+V^2)}[\nu_{1}^2+
\beta \nu_{1}^4]
-\frac{m^2}{\sigma^2}-\frac{eA_{0}U}{\sigma^2(U\Delta+V^2)}
[(E-j\omega)\nonumber\\
&+&\beta (E-j\omega)^3-eA_{0}-\beta eA_{0}(E-j\omega)^2],\nonumber\\
Z_{23}&=&\frac{\Delta}{\sigma^2(U\Delta+V^2)}[J+\beta J^3]\nu_{1},\nonumber\\
Z_{30}&=&\frac{(U\Delta-\Delta^2)}{(U\Delta+V^2)^2}[\nu_{1}+\beta \nu_{1}^3](E-j\omega)+ \frac{V}{\sigma^2(U\Delta+V^2)}[J^2+\beta J^4]-\frac{m^2V}{(U\Delta+V^2)}-\frac{eA_{0}(U\Delta-\Delta^2)}{(U\Delta
+V^2)^2}[\nu_{1}+\beta \nu_{1}^3],\nonumber
\end{eqnarray}
\begin{eqnarray}
Z_{31}&=&\frac{\Delta}{(U\Delta+V^2)}[\nu_{1}+\beta \nu_{1}^3]W_{1},\nonumber\\
Z_{32}&=&\frac{V}{\sigma^2(U\Delta+V^2)}[J+\beta J^3](E-j\omega)+
\frac{\Delta}{\sigma^2(U\Delta+V^2)}[\nu_{1}+\beta \nu_{1}^3]J,\nonumber\\
Z_{33}&=&\frac{(U\Delta-\Delta^2)}{(U\Delta+V^2)}[(E-j\omega)^2
+\beta (E-j\omega)^4-eA_{0}(E-j\omega)-\beta eA_{0}(E-j\omega)^3]
-\frac{\Delta}{(U\Delta+V^2)}[W_{1}^2+\beta W_{1}^4]
\nonumber\\
&-&\frac{\Delta}{\sigma^2(U\Delta+V^2)}[J^2+\beta J^4]
-\frac{m^2 \Delta}{(U\Delta+V^2)}
-\frac{eA_{0}(U\Delta-\Delta^2)}{(U\Delta+V^2)}[(E-j\omega)
+\beta (E-j\omega)^3-eA_{0}(E-j\omega)^2],\nonumber
\end{eqnarray}
whereas $J=\partial_{\phi}S_{0}$, $W_{1}=\partial_{r}{S_{0}}$, $\nu_{1}=\partial_{\theta}{S_{0}}$.
For the purpose of non-trivial solution, we set the determinant
of $\textbf{Z}$ equals to the and attain the solution in the form
\begin{eqnarray}\label{a1}
Im W^{\pm}&=&\pm \int\sqrt{\frac{(E-j\omega-eA_{0})^{2}
+X_{1}\Big[1+\beta\frac{X_{2}}{X_{1}}\Big]}{\frac{(U\Delta+V^2)\Delta}{U}}}dr,\nonumber\\
&=&\pm i\pi\frac{(E-j\omega-eA_{0})+[1+\beta\Xi]}{2\kappa(r_{+})},
\end{eqnarray}
where, $\Xi$ is a arbitrary parameter and $X_{1}$, $X_{2}$ are defined as
\begin{eqnarray}
X_{1}&=&\frac{V}{(U\Delta+V^2)\Delta}[(E-j\omega)
-eA_{0}]\nu_{1}+\frac{\nu_{1}^2}{(U\Delta+V^2)}-\frac{m^2}{\Delta},\nonumber\\
X_{2}&=&\frac{U}{\Delta(U\Delta+V^2)}[(E-j\omega)^4
-2eA_{0}(E-j\omega)^3+(eA_{0})^2(E-j\omega)^2]-\frac{\nu_{1}^4}{(U\Delta+V^2)}-W_{1}^4\nonumber\\
&+&\frac{V}{\Delta\sigma^2(U\Delta+V^2)}[(E-j\omega)^3-eA_{0}(E-j\omega)^2]\nu_{1}
.\nonumber
\end{eqnarray}
The tunneling probability \cite{tp} of charged boson particles is derived as
\begin{equation}
\Gamma\simeq\exp\Big[{-2\pi}\frac{(E-j\omega-eA_{0})}
{\kappa(r_{+})}\Big]\Big[1+\beta\Xi\Big].
\end{equation}
The corrected temperature of parameterized BH with rotation parameter can be computed in the form
\begin{equation}
T{'}_{H}=\frac{2M r_{+}^4+3\eta r_{+}^2+a^2 (\eta-2M r_{+}^2)}{4\pi r_{+}^{2}\Big(r_{+}^2 +a^2\Big)^2}\Big[1-\beta \Xi\Big].\label{cht}
\end{equation}
The modified Hawking temperature $T'_{H}$ of BH depends upon rotation parameter
$a$, deformation parameter $\eta$, mass $M$, radius horizon $r_{+}$ as well as quantum gravity parameter $\beta$.
When the correction parameter $\beta=0$, we get the original temperature of Eq. (\ref{2}) as well as by putting $\eta=0=a$, we obtain the Hawking temperature of Schwarzschild BH $T_{Sch}=\frac{1}{8\pi M}$ at $r_+\thickapprox 2M$.

\section{Graphical Explanation of Corrected Temperature}
This section demonstrates the graphical description of corrected temperature $T'_{H}$
versus $r_{+}$. We analyze the physical
value of the plots with the impacts of deformation parameter $\eta$, spin parameter $a$ and GUP parameter $\beta$.

\begin{center}
\includegraphics[width=7cm]{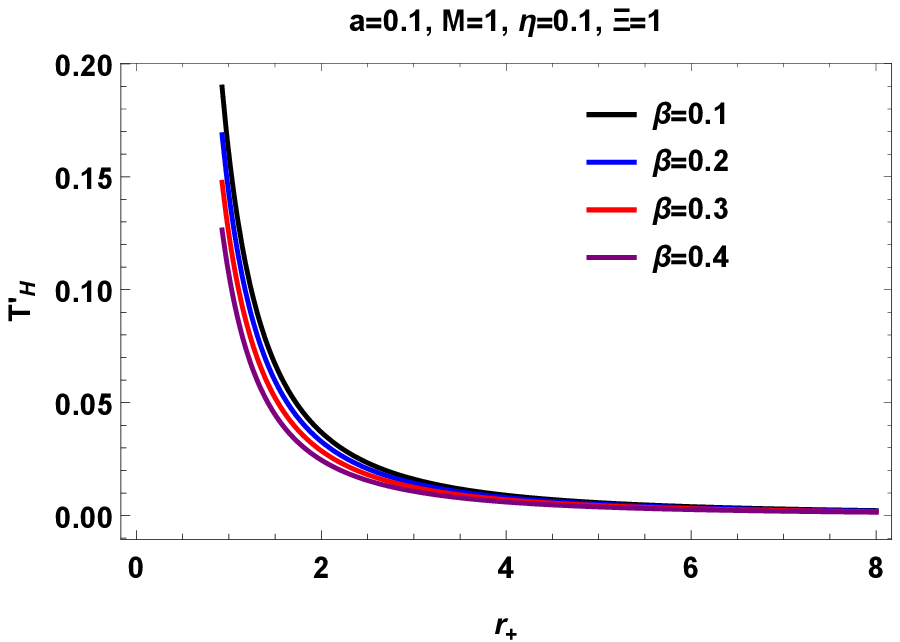}\includegraphics[width=7cm]{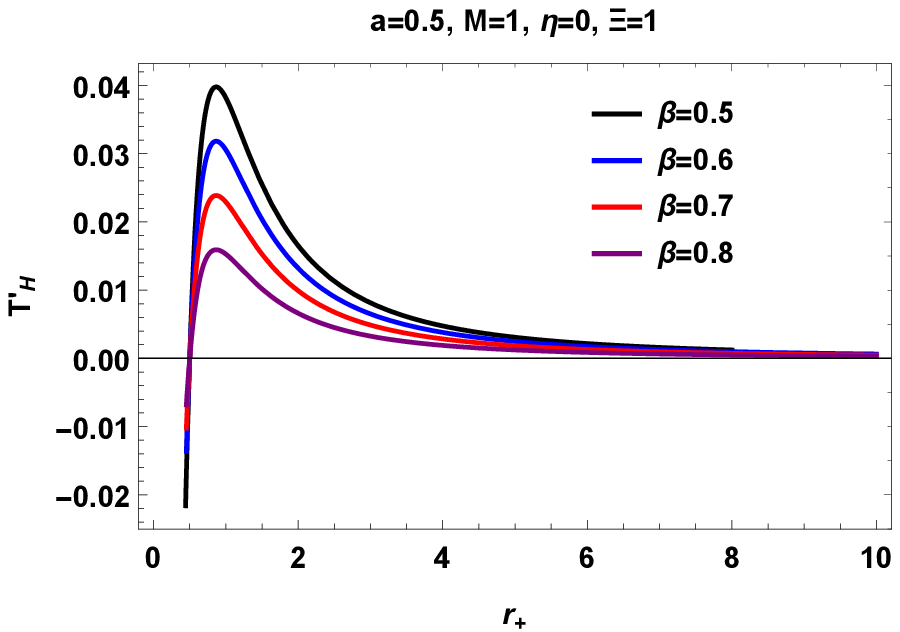}\\
{\text{Fig. 2}: Corrected Temperature via horizon for $M=1=\Xi$. Left $a=0.1,~\eta=0.1$, $\beta=0.1$ (black), $0.2$ (blue), $0.3$ (red), and $0.4$ (purple).} Right $a=0.5,~\eta=0$, $\beta=0.5$ (black), $0.6$ (blue), $0.7$ (red), and $0.8$ (purple).
\end{center}
The left plot in \textbf{Fig. 2} represents that the $T'_{H}$ exponentially decreases for rising horizon in the region $0\leq r_{+}\leq8$ for different variations of $\beta$. For rising values of $\beta$ the $T'_{H}$ decreases.

The right plot depicts that afterwards gaining a peak height the $T'_{H}$ suddenly falls down in a decreasing way and reaches an asymptotically flat state until $r_{+}\rightarrow\infty$. In the absence of deformation parameter, the $T'_{H}$ also decreases within the increasing variations of $\beta$ in $0\leq r_{+}\leq10$.

As stated by Hawking's scenario with the increasing temperature more
radiation emit and the radius of BH decreases. In \textbf{Fig. 2}, one can examine that the temperature is maximum at minimum horizon. This physical strategy demonstrates the stable state of BH.
\begin{center}
\includegraphics[width=7cm]{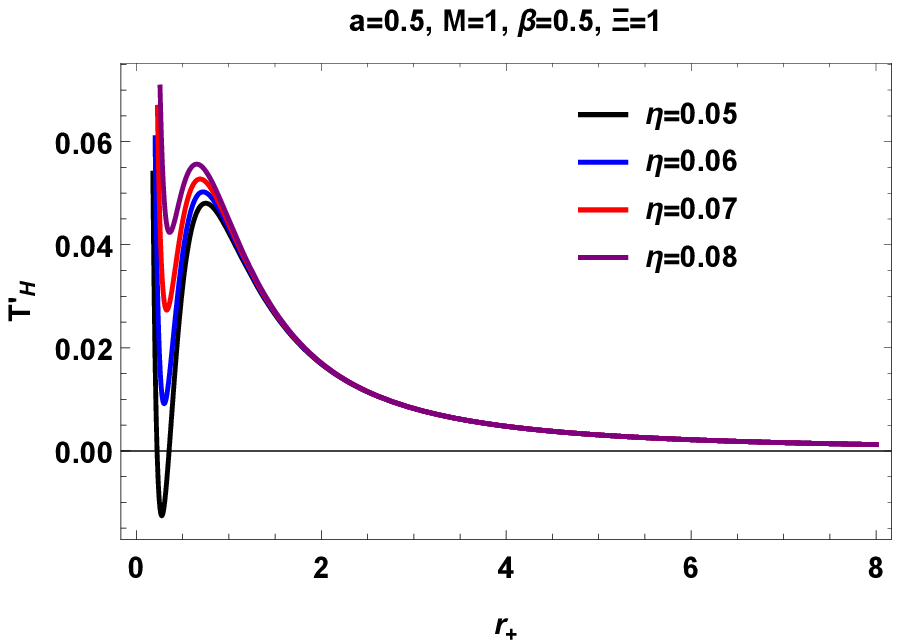}\includegraphics[width=7cm]{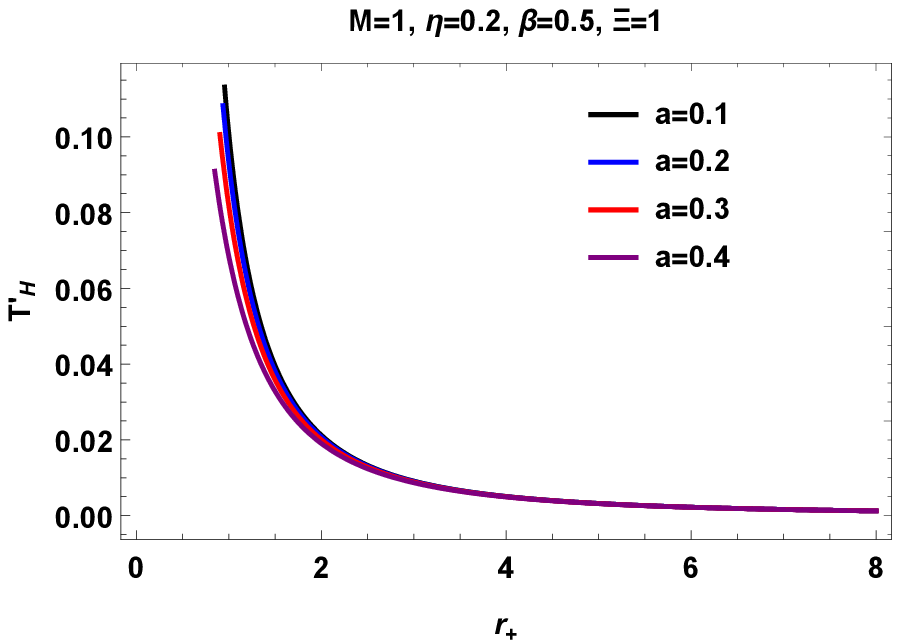}\\
{\text{Fig. 3}: Corrected Temperature w. r. t horizon for $M=1=\Xi$. Left $a=0.5=\beta$, $\eta=0.05$ (black), $\eta=0.06$ (blue), $\eta=0.07$ (red), and $\eta=0.08$ (purple). Right $\beta=0.5=\eta$, $a=0.1$ (black), $a=0.2$ (blue), $a=0.3$ (red), and $a=0.4$ (purple)}.
\end{center}
The left plot in \textbf{Fig. 3} indicates, the corrected temperature slowly decreases with distinct variations of $\eta$ in $0\leq r_{+}\leq8$ and after a minima it rises to reach at a maxima and then again gradually decrease to shows an asymptotically flat description until $r\rightarrow\infty$. The plot represents the two critical values $r^{min}_{+}$ and $r^{max}_{+}$. The BH assures its stability for $r^{min}_{+}\leq r_{+}\leq r^{max}_{+}$ by getting an asymptotically flat state upon $r_{+}\rightarrow\infty$. In the existence of correction parameter with rising $\eta$, the $T'_{H}$ also rises.

The right plot demonstrate the conduct of $T'_{H}$ for distinct values of $a$ in $0\leq r_{+}\leq8$.

Moreover, it is an admitted fact in quantum gravity that the quantum corrections are a cause of reduction in rising temperature. So, this phenomenon can be clearly observed in plots with the presence/absence of gravity parameter.

\section{Entropy Corrections for Parameterized Black Hole within Spin Parameter}
This section studies the Bekenstein-Hawking entropy associated with correction factor in the theory of
quantum loop expansion. The logarithmic expression appears as an additional loop into the standard entropy of BH. The entropy of BH has a very significant role in primordial BHs \cite{A1}. It established a least mass of the primordial BHs so that the small BHs have sufficient opportunity
to vanish completely in the current form.
We investigate the entropy corrections for parameterized BH with rotation parameter. The entropy corrections via null
geodesic technique has been studied by Banerjee and Majhi \cite{A2}-\cite{A4}. We investigate the entropy corrections for parameterized BH with rotation parameter by utilizing the formula of Bekenstein-Hawking entropy for $1^{st}$ order corrections \cite{A5}.
By using the expressions of $T'_{H}$ and standard entropy $\mathbb{S}_{T}$, we calculate the logarithmic entropy corrections for parameterized BH with rotation parameter through the given formula
\begin{equation}
\mathbb{S}_{P}=\mathbb{S}_{T}-\frac{1}{2}\ln\Big|T_H^2, \mathbb{S}_{T}\Big|+...~.\label{vv}
\end{equation}
The standard entropy for parameterized BH with rotation parameter can be derived through the given formula
\begin{equation}
\mathbb{S}_{T}=\frac{A_P}{4},
\end{equation}
where
\begin{eqnarray}
A_P&=&\int_{0}^{2\pi}\int_{0}^{\pi}\sqrt{g_{\theta\theta}g_{\phi\phi}}d\theta d\phi,\nonumber\\
&=&\pi\left[\frac{2(r_{+}^2+a^2\Big)^2
+a^2\Big(2Mr_{+}+\eta r^{-1}\Big)}{\Big(r_{+}^2+a^2\Big)}\right].
\end{eqnarray}
The standard entropy for parameterized BH with rotation parameter can be calculated in the form
\begin{equation}
\mathbb{S}_{T}=\pi\left[\frac{2(r_{+}^2+a^2\Big)^2
+a^2\Big(2Mr_{+}+\eta r^{-1}\Big)}{4\Big(r_{+}^2+a^2\Big)}\right].\label{v1}
\end{equation}
After substituting the values from (\ref{v1}) into Eq. (\ref{vv}), we obtain the entropy corrections as
\begin{eqnarray}
\mathbb{S}_{P}&=&\pi\left[\frac{2(r_{+}^2+a^2\Big)^2
+a^2\Big(2Mr_{+}+\eta r^{-1}\Big)}{4\Big(r_{+}^2+a^2\Big)}\right]\nonumber\\
&-&\frac{1}{2}\ln\left|\frac{\left[\Big\{2M r_{+}^4+3\eta r_{+}^2+a^2 (\eta-2M r_{+}^2)\Big\}\Big\{1-\beta\Xi\Big\}\right]^2\Big[2\Big(r_{+}^2+a^2\Big)^2
+a^2\Big(2Mr_{+}+\eta r^{-1}\Big)\Big]}{64\pi r^{4}_{+}\left(r^2_++a^2\right)^5}\right|+...,\label{b2}
\end{eqnarray}

The Eq. (\ref{b2}) shows the corrected entropy for parameterized BH with rotation parameter.

\section{Corrected Entropy Analysis}

The graphical analysis of entropy corrections $S_{P}$ versus horizon $r_+$ for parameterized BH with rotation parameter has been analyzed in this section. We analyze the impacts of $\beta$ and $\eta$ on corrected entropy $S_{P}$ via graphs.
\begin{center}
\includegraphics[width=8cm]{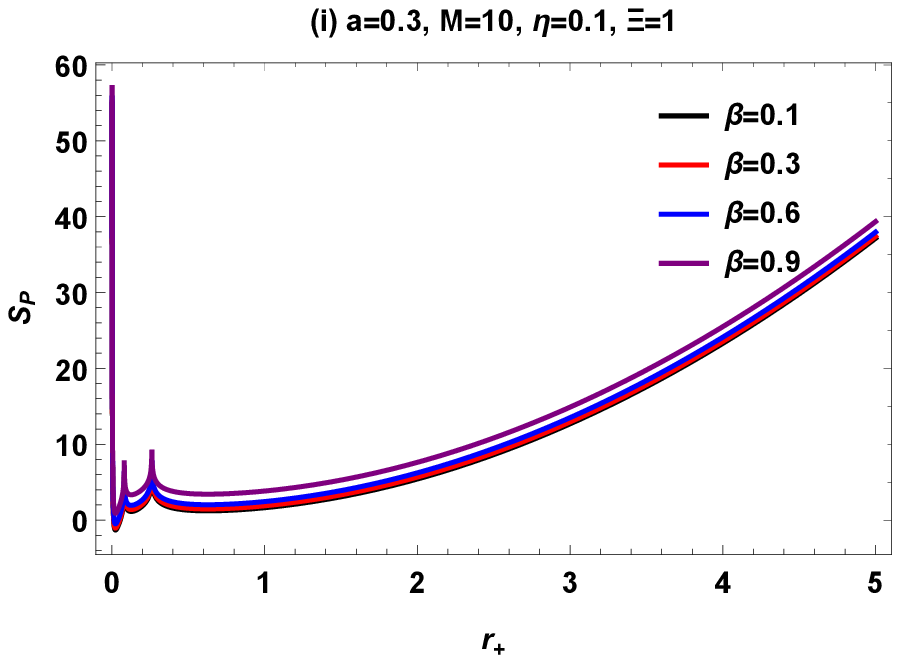}\includegraphics[width=8cm]{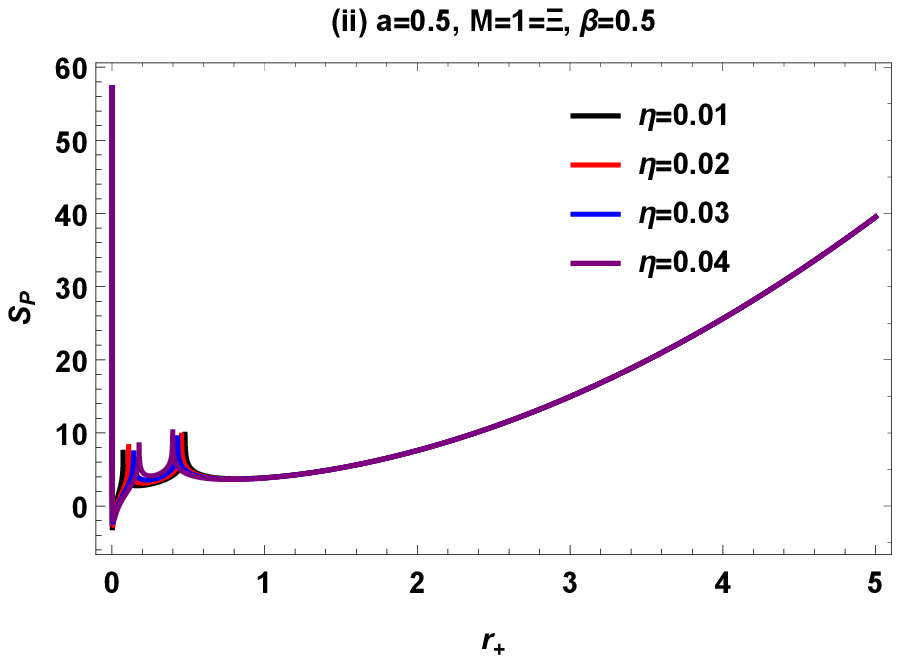}\\
{Figure 4: Corrected Entropy $S_{P}$ via horizon $r_{+}$ for $\Xi=1$. Left $a=0.3, \eta=0.1$, $M=10$, $\beta=0.1$ (black), $0.3$ (red), $0.6$ (blue) and $0.8$ (purple). Right $a=0.5, \beta=0.5, M=1$, $\eta=0.01$ (black), $0.02$ (red), $0.03$ (blue) and $0.04$ (purple)}.
\end{center}

The left plot in \textbf{Fig. 4} depicts the graphical conduct of $S_{P}$ for changing values of $\beta$ in the region $0\leq r_{+}\leq 5$. The corrected entropy slowly decreases and after showing some variations the
$S_{P}$ continuously increases for rising values of horizon. This conduct with positive entropy shows the stable form of parameterized BH. The $S_{P}$ increases with the rising values of correction parameter.

The right plot in \textbf{Fig. 4} gives the behavior of $S_{P}$ versus $ r_{+}$ for different values of deformation parameter $\eta$ in the region $0\leq r_{+}\leq 5$.
One can see
from the plot that with the rising values of deformation parameter $\eta$ the $S_{P}$ also increases. It is also observable that at most of the points sharp edge appears for varying values of deformation parameter that reduces the stability of entropy.

It can be conclude from the observation of both plots, the BH entropy is more stable for varying values of correction parameter as compared to deformation parameter.

\section{Results and Discussion}

In this research work, we have investigated the solution of parameterized BH with spin parameter by using the Newman-Janis strategy. By setting the rotation coupling
$(a\rightarrow 0)$ in the Eq. (\ref{ds}), we obtained the metric of parameterized BH free off rotation parameter.
The solution of parameterized BH within Newman-Janis approach is actually in contrast from the BH solution in general relativity.
Furthermore, we have studied the thermodynamics for parameterized BH with rotation parameter. The temperature is dependent on the mass $M$,
deformation parameter $\eta$, radius of BH $r_+$ as well as spin parameter $a$. We recover the temperature for parameterized BH without rotation parameter for $a=0$.
 We have analyzed the physical importance of temperature versus horizon under the effects of deformation and rotation parameter.
The $T_{H}$ rises with the rising values of deformation parameter $\eta$.

The $T_{H}$ constantly decreases and after a minima it rises to reach at a maxima and then again gradually decrease to shows an asymptotically flat description until $r\rightarrow\infty$ for changing values of deformation parameter. There can be seen two critical values $r_{+}^{min}$ and $r_{+}^{max}$ in the plot where the BH is stable for $r_{+}^{min}\leq r_{+}\leq r_{+}^{max}$. For the rising values of $a$, the $T_{H}$ shows a decreasing conduct. The stability of BH can be observed by Hawking's scenario with the increasing temperature more
radiation emit and the radius of BH decreases.
Furthermore, we have studied the corrected temperature for parameterized BH with rotation parameter accompanying by GUP effects.
When the correction parameter $\beta=0$, we get the temperature of Eq. (\ref{2}) as well as by putting $\eta=0=a$, we obtain the Hawking temperature of Schwarzschild BH $T_{Sch}=\frac{1}{8\pi M}$ at $r_+\thickapprox2M$. We have also studied the graphical evaluation of corrected temperature with respect to horizon and graphically discussed the
stable condition of parameterized BH with rotation parameter under the impact of quantum gravity.
The $T'_{H}$ exponentially decreases for rising horizon in the region $0\leq r_{+}\leq8$ for different variations of $\beta$. For rising values of $\beta$ the $T'_{H}$ decreases.
In the absence of deformation parameter, the $T'_{H}$ also decreases with the rising variations of $\beta$ in the region $0\leq r_{+}\leq10$.
 In \textbf{Fig. 2}, one can examined that the temperature is maximum at minimum horizon. This physical strategy demonstrates the stable state of BH.
 In the existence of gravity parameter with rising $\eta$, the $T'_{H}$ also rises.

The conduct of $T'_{H}$ for distinct values of $\eta$ in $0\leq r_{+}\leq8$ demonstrates the two critical values $r^{min}_{+}$ and $r^{max}_{+}$. The BH assures its stability for $r^{min}_{+}\leq r_{+}\leq r^{max}_{+}$ by getting an asymptotically flat state upon $r_{+}\rightarrow\infty$.

Moreover, it is an admitted fact in quantum gravity that the quantum corrections are a cause of reduction in rising temperature. We have concluded that the temperature
in {\bf Fig. 1} (absence of quantum gravity parameter) is higher than in {\bf Fig. 3} (presence of quantum gravity parameter), so, this phenomenon can be clearly observed in plots with the presence/absence of gravity parameter.

At last, the corrected entropy for parameterized BH with rotation parameter is calculated. By neglecting the gravity parameter effects, we attained the standard entropy of parameterized BH with rotation parameter.
The corrected entropy slowly decreases and after some variations the $S_{P}$ continuously increases for rising values of horizon. This conduct shows the stable form of parameterized BH. The $S_{P}$ increases with the increasing values of correction parameter.

For different values of deformation parameter $\eta$ in the region $0\leq r_{+}\leq 5$, one can observed that with the rising values of deformation parameter $\eta$ the $S_{P}$ also increases. It is also observable that at most of the points sharp edge appears for varying values of deformation parameter that reduces the stability of entropy.

It can be concluded from the observation of both plots of corrected entropy, the BH entropy is more stable for varying values of correction parameter as compared to deformation parameter.

\section{Appendix A}
We apply the WKB approximation in Lagrangian field equation and collect a set of field equations in the form
\begin{eqnarray}
&&\frac{U\Delta}{(U\Delta+V^2)}[c_{1}(\partial_{0}I_{0})(\partial_{1}I_{0})+\beta c_{1}
(\partial_{0}I_{0})^{3}(\partial_{1}I_{0})-c_{0}(\partial_{1}I_{0})^{2}
-\beta c_{0}(\partial_{1}I_{0})^4\nonumber\\
&+&c_{1}eA_{0}(\partial_{1}I_{0})+c_{1}\beta eA_{0}(\partial_{0}I_{0})^{2}(\partial_{1}I_{0})]
-\frac{V\Delta}{(U\Delta+V^2)}[c_{3}(\partial_{1}I_{0})^2+\beta c_{3}(\partial_{1}I_{0})^4\nonumber\\
&&-c_{1}(\partial_{1}I_{0})(\partial_{3}I_{0})-\beta c_{1}(\partial_{1}I_{0})(\partial_{3}I_{0})^2]
+\frac{U}{\sigma^2(U\Delta+V^2)}[c_{2}(\partial_{0}I_{0})(\partial_{2}I_{0})
+\beta c_{2}\nonumber\\
&&(\partial_{0}I_{0})^3(\partial_{2}I_{0})
-c_{0}(\partial_{2}I_{0})^2-\beta c_{0}(\partial_{2}I_{0})^4+c_{2}eA_{0}(\partial_{2}I_{0})+c_{2}eA_{0}\beta(\partial_{0}I_{0})^{2}(\partial_{1}I_{0})]\nonumber\\
&+&\frac{U\Delta}{(U\Delta+V^2)^2}[c_{3}(\partial_{0}I_{0})(\partial_{3}S_{0})
+\beta c_{3}(\partial_{0}I_{0})^{3}(\partial_{3}I_{0})-c_{0}(\partial_{3}I_{0})^{2}
-\beta c_{0}(\partial_{3}I_{0})^4\nonumber\\
&+&c_{3}eA_{0}(\partial_{3}I_{0})+c_{3}eA_{0}(\partial_{0}I_{0})^{2}(\partial_{3}I_{0})]-m^2\frac{U c_{0}-{V c_{3}}}{(U\Delta+V^2)}=0,\label{E1}\\
&-&\frac{U\Delta}{(U\Delta+V^2)}[c_{1}(\partial_{0}I_{0})^2+\beta c_{1}
(\partial_{0}I_{0})^4-c_{0}(\partial_{0}I_{0})(\partial_{1}I_{0})-\beta c_{0}(\partial_{0}I_{0})(\partial_{1}I_{0})^{3}\nonumber\\
&+&c_{1}eA_{0}(\partial_{0}I_{0})+\beta c_{1}eA_{0}(\partial_{0}I_{0})^3]+\frac{V\Delta}{(U\Delta+V^2)}
[c_{3}(\partial_{0}I_{0})(\partial_{1}I_{0})+\beta c_{3}
(\partial_{0}I_{0})\nonumber\\
&&(\partial_{1}I_{0})^3-c_{1}(\partial_{0}I_{0})(\partial_{3}I_{0})-\beta c_{1}(\partial_{0}I_{0})(\partial_{3}I_{0})^{3}]+\frac{\Delta}{\sigma^2}[c_{2}(\partial_{1}I_{0})(\partial_{2}I_{0})+\beta c_{2}\nonumber\\
&&(\partial_{1}I_{0})(\partial_{2}I_{0})^3-c_{1}(\partial_{2}I_{0})^{2}-\beta c_{1}(\partial_{2}I_{0})^{4}]+\frac{\Delta}{(U\Delta+V^2)}[c_{3}(\partial_{1}I_{0})(\partial_{3}I_{0})\nonumber\\
&+&\beta c_{3}(\partial_{1}I_{0})(\partial_{3}I_{0})^3-c_{1}(\partial_{3}I_{0})^2-\beta c_{1} (\partial_{3}I_{0})^{4}]-m^2 \Delta c_{1}
+\frac{eA_{0}U\Delta}{(U\Delta+V^2)}[c_{1}\nonumber\\
&&(\partial_{0}I_{0})+\beta c_{1}(\partial_{0}I_{0})^3
-c_{0}(\partial_{1}I_{0})-\beta c_{0}(\partial_{1}I_{0})^3+eA_{0}c_{1}+\beta c_{1}eA_{0}(\partial_{0}I_{0})^{2})]\nonumber\\
&+&\frac{eA_{0}V\Delta}{(U\Delta+V^2)}[c_{3}(\partial_{1}I_{0})+\beta c_{3}(\partial_{1}I_{0})^3
-c_{1}(\partial_{3}I_{0})-\beta c_{1}(\partial_{1}I_{0})^3]=0,\label{E2}\\
&&\frac{U}{\sigma^2(U\Delta+V^2)}[c_{2}(\partial_{0}I_{0})^2+\beta c_{2}
(\partial_{0}I_{0})^{4}-c_{0}(\partial_{0}I_{0})(\partial_{2}I_{0})
-\beta c_{0}(\partial_{0}I_{0})(\partial_{2}I_{0})^3\nonumber\\
&+&c_{2}eA_{0}(\partial_{0}I_{0})+\beta c_{2}eA_{0}(\partial_{0}I_{0})^{3}]
+\frac{\Delta}{\sigma^2}[c_{2}(\partial_{1}I_{0})^2+\beta c_{2}
(\partial_{1}I_{0})^{4}-c_{1}(\partial_{1}I_{0})\nonumber\\
&&(\partial_{2}I_{0})
-\beta c_{1}(\partial_{1}I_{0})(\partial_{2}I_{0})^3]-\frac{V}{\sigma^2(U\Delta
+V^2)}[c_{2}(\partial_{0}I_{0})(\partial_{3}I_{0})+\beta c_{2}
(\partial_{0}I_{0})^{3}\nonumber\\
&&(\partial_{3}I_{0})-c_{0}(\partial_{0}I_{0})(\partial_{3}I_{0})
-\beta c_{0}(\partial_{0}I_{0})^3 (\partial_{3}I_{0})+c_{2}eA_{0}(\partial_{3}I_{0})
+\beta c_{2}eA_{0}\nonumber\\
&&(\partial_{3}I_{0})^{3}]
+\frac{\Delta}{\sigma^2(U\Delta+V^2)}[c_{3}(\partial_{2}I_{0})(\partial_{3}I_{0})+\beta c_{3}
(\partial_{2}I_{0})^{3}(\partial_{3}I_{0})-c_{2}(\partial_{3}I_{0})^2\nonumber\\
&-&\beta c_{2}(\partial_{3}I_{0})^4]
-\frac{m^2 c_{2}}{\sigma^2}+\frac{eA_{0}U}{\sigma^2(U\Delta+V^2)}[c_{2}(\partial_{0}I_{0})+\beta c_{2}(\partial_{0}I_{0})^3-c_{0}(\partial_{2}I_{0})\nonumber\\
&-&\beta c_{0}(\partial_{2}I_{0})^3+c_{2}eA_{0}+c_{2}\beta eA_{0}(\partial_{0}I_{0})^2]=0,\label{E3}
\end{eqnarray}
\begin{eqnarray}
&&\frac{(U\Delta)-\Delta^2}{(U\Delta+V^2)^2}[c_{3}(\partial_{0}I_{0})^2+\beta c_{3}
(\partial_{0}I_{0})^4-c_{0}(\partial_{0}I_{0})(\partial_{3}I_{0})-\beta c_{0}(\partial_{0}I_{0})(\partial_{3}I_{0})^{3}\nonumber\\
&+&{eA_{0}c_3}(\partial_{0}I_{0})
+\beta c_{3}eA_{0}(\partial_{0}I_{0})^{3}]
-\frac{U}{\sigma^2(U\Delta+V^2)}[c_{3}(\partial_{1}I_{0})^2+\beta c_{3}
(\partial_{1}I_{0})^{4}\nonumber\\
&-&c_{1}(\partial_{1}I_{0})(\partial_{3}I_{0})
-\beta c_{1}(\partial_{1}I_{0})(\partial_{3}I_{0})^3]-\frac{V}{\sigma^2(U\Delta+V^2)}
[c_{2}(\partial_{0}I_{0})(\partial_{2}I_{0})\nonumber\\
&+&\beta c_{2}
(\partial_{0}I_{0})^3(\partial_{2}I_{0})-c_{0}(\partial_{2}I_{0})^{2}
+\beta c_{0}(\partial_{2}I_{0})^4+{eA_{0}c_2}(\partial_{2}I_{0})
+\beta c_{2}eA_{0}\nonumber\\
&&(\partial_{0}I_{0})^{2}(\partial_{2}I_{0})]-\frac{eA_{0}\Delta}{\sigma^2(U\Delta+V^2)}
[c_{3}(\partial_{2}I_{0})^2+\beta c_{3}(\partial_{2}I_{0})^4-c_{2}(\partial_{2}I_{0})(\partial_{3}I_{0})\nonumber\\
&-&\beta c_{2}(\partial_{0}I_{0})(\partial_{3}I_{0})^{3}]
-\frac{m^2 (Vc_{0}-\Delta c_{3})}{(U\Delta+V^2)}
+\frac{U\Delta eA_{0}-\Delta^2}{(U\Delta+V^2)^2}[c_{3}(\partial_{0}I_{0})+\beta c_{3}\nonumber\\
&&(\partial_{0}I_{0})^3
-c_{0}(\partial_{3}I_{0})-\beta c_{0}(\partial_{3}I_{0})^3+c_{3}eA_{0}
+\beta eA_{0}(\partial_{0}I_{0})^2]=0.\label{E4}
\end{eqnarray}

 \section*{Appendix B}
 We have made the assumption that the Lagrangian gravity equation exists in order to see the bosonic tunneling. Additionally, we generated a set of equations using the WKB approximation to the Lagrangian gravity equation. So, we have used the variable separation action to obtain the solutions in matrix form. The equations (\ref{E1})-(\ref{E4}) of general matrix form can be expressed as
\[\begin{pmatrix}
0\\
0\\
0\\
0
\end{pmatrix}=\begin{pmatrix}
Z_{00}&Z_{01}&Z_{02}&Z_{03}\\
Z_{10}&Z_{11}&Z_{12}&Z_{13}\\
Z_{20}&Z_{21}&Z_{22}&Z_{23}\\
Z_{30}&Z_{31}&Z_{32}&Z_{33}\\
\end{pmatrix}\begin{pmatrix}
c_{0}\\
c_{1}\\
c_{2}\\
c_{3}\\
\end{pmatrix}\]

\end{document}